\documentclass[11pt,a4paper]{article}
\usepackage[hyperref]{emnlp-ijcnlp-2019}
\usepackage{times}
\usepackage{latexsym}

\usepackage{url}

\aclfinalcopy % Uncomment this line for the final submission

\usepackage{enumerate}
\usepackage[subrefformat=parens]{subcaption}
\usepackage{comment}
\usepackage{graphicx}
\usepackage{multirow}
\usepackage[whole]{bxcjkjatype}

\title{Character 3-gram Mover's Distance: An Effective Method for Detecting Near-duplicate Japanese-language Recipes}

\author{Masaki Oguni \\
  University of Tsukuba\\
  {\tt m-oguni@klis.tsukuba.ac.jp} \\\And
  \textbf{Yohei Seki} \\
  University of Tsukuba \\
  {\tt yohei@slis.tsukuba.ac.jp} \\\AND
  Yu Hirate \\
  Rakuten Inc.\\
  {\tt yu.hirate@rakuten.com} \\}

\date{}

\begin{document}
\maketitle
\begin{abstract}
%In websites that collect user-generated recipes, recipes are often posted that have a major component, such as the cooking instructions, that is very similar to those in other recipes.
In user-generated recipe websites, users post their-original recipes. Some recipes, however, are very similar in major components such as the cooking instructions to other recipes.
We refer to such recipes as ``near-duplicate recipes''.
In this study, we propose a method that extends the ``Word Mover's Distance'', which calculates distances between texts based on word embedding, to character 3-gram embedding.
Using a corpus of over 1.21 million recipes, we learned the word embedding and the character 3-gram embedding by using a Skip-Gram model with negative sampling and fastText to extract candidate pairs of near-duplicate recipes.
We then annotated these candidates and evaluated the proposed method against a comparison method.
Our results demonstrated that near-duplicate recipes that were not detected by the comparison method were successfully detected by the proposed method.
\end{abstract}

\section{Introduction}
Many people access and search for recipes via recipe websites.
On several recipe websites, users can post recipes they have created themselves.
We refer to these recipe websites as ``user-generated recipe websites''.
%Rakuten Recipe\footnote{https://recipe.rakuten.co.jp/} is one of the best-known user-generated recipe websites in Japan, posting over 1.68 million recipes by August 2019.

On user-generated recipe websites, recipes are often posted that have a major component, such as the cooking instructions, that is very similar to those in other recipes.
We refer to these recipes as ``near-duplicate recipes'' and a recipe considered to be the source recipe for near-duplicates as their ``original recipe''.
Finally, we refer to a recipe pair comprising a near-duplicate recipe and its original recipe as a ``near-duplicate recipe pair''. 

Kusner et al. \shortcite{kusner:2015} proposed a suitable method for similar-document searching called the ``Word Mover's Distance'' (WMD), which calculates distances between texts in terms of word embedding.
Oguni et al. \shortcite{oguni:2017} reported that the character 3-gram is suited to detecting Japanese-language near-duplicate recipes.
Following these ideas, we propose a method that extends the WMD approach to character 3-gram embedding.
We refer to this method as the ``Character 3-gram Mover's Distance'' method.

The contributions of our paper are that:

\textbullet we propose a near-duplicate detection method suitable for Japanese-language recipes

\textbullet we investigate differences in the performance of two embedding-based learning algorithms for Japanese-language near-duplicate recipe detection.

\begin{figure*}[htb]
\centering
  \begin{minipage}[b]{.55\linewidth}
    \centering
    \includegraphics[width=8.5cm]{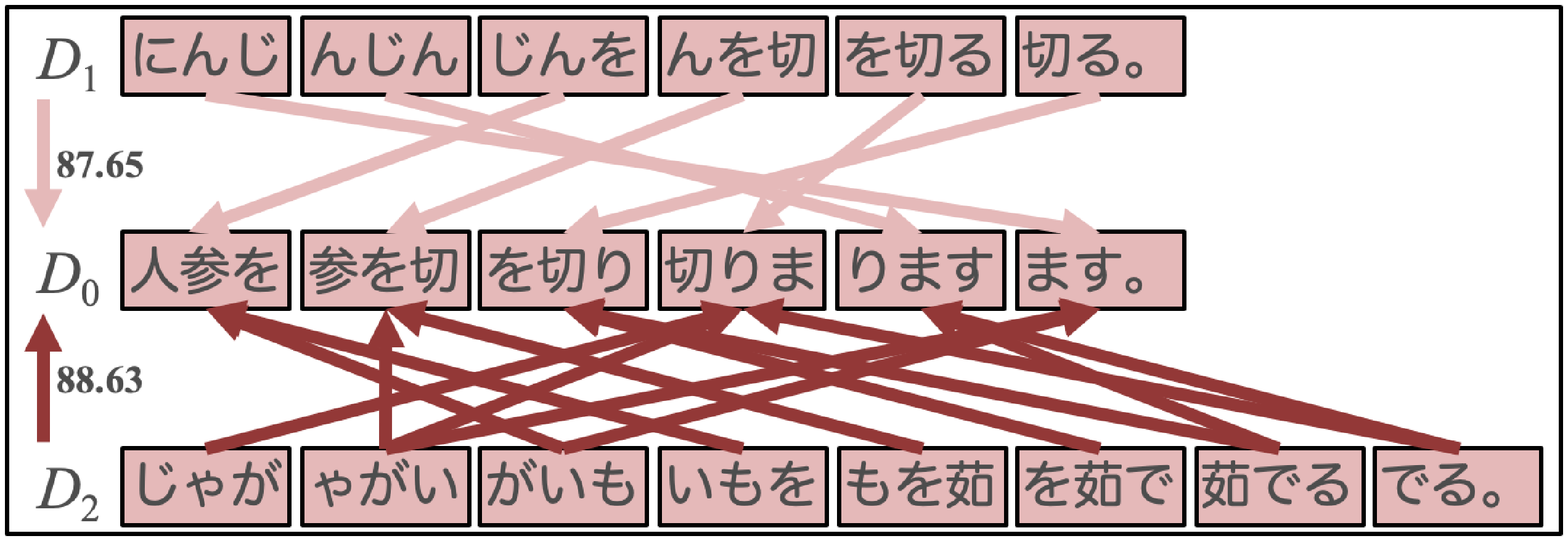}
    \subcaption{Proposed method: \\ sentences divided by character 3-grams}%\label{method-a}
  \end{minipage}
  \begin{minipage}[b]{.44\linewidth}
    \centering
    \includegraphics[width=4.5cm]{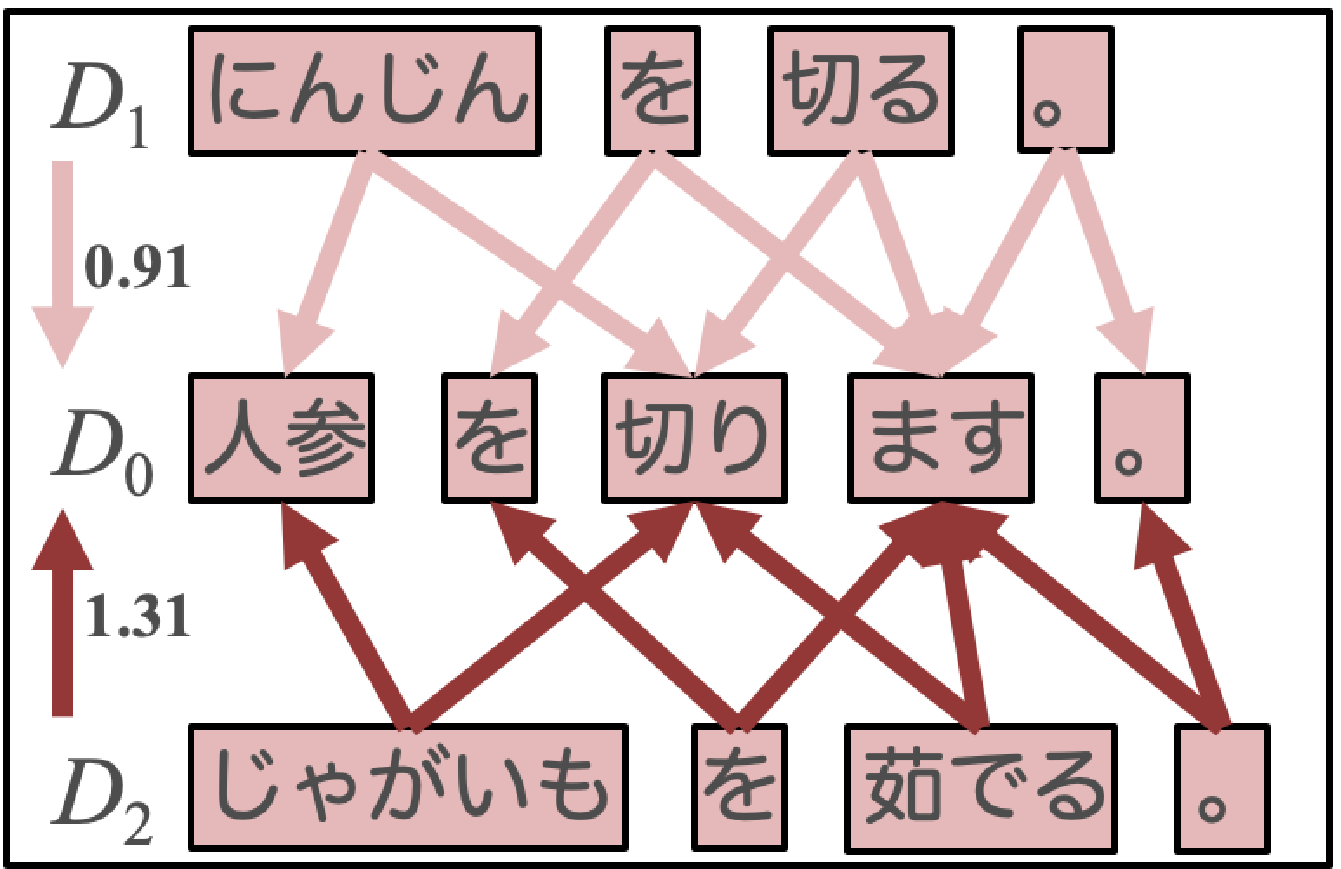}
    \subcaption{Comparison method: \\ sentences divided by words}%\label{method-b}
  \end{minipage}
  \caption{Calculation methods for distances between cooking-instruction texts}
  \label{method}
\end{figure*}

\section{Related work}

Oguni et al. \shortcite{oguni:2017} showed that the character 3-gram is suited to detecting Japanese-language near-duplicate recipes and utilized the Jaccard index between character 3-gram sets as a measure of cooking-instruction text similarity. However, in recipes, there is text that can be paraphrased and ingredients that can have variations in naming. 
Furthermore, because we are using a user-generated recipe dataset as the experimental dataset, recipes may have typographical errors and omissions.
From these considerations, we cannot expect to detect near-duplicate recipes simply by comparing character 3-gram values between recipes.
%From these considerations, we cannot expect to detect near-duplicate recipes by comparing character 3-gram values between recipes.

Kusner et al. \shortcite{kusner:2015} proposed the WMD method as being suited to searching for similar documents.
WMD measures the dissimilarity between two texts as the minimum distance that the embedded words of one document need to travel to reach the embedded words of the other document.

Several methods for extracting textual features from texts have been proposed \cite{kiros:2015, mekala:2017, arora:2017, logeswaran:2018}.
In these methods, the features of one document are concatenated into a single feature vector.
Therefore, information about the replacement or rewriting of words will be lost when the similarity between cooking-instruction texts is calculated.

Following these ideas, we propose a method that extends the WMD approach to character 3-gram embedding.
In this way, our method should be robust against typographical errors and omissions when detecting near-duplicate recipes.

\section{Proposed method}

%We propose a method that extends the WMD approach to character 3-gram embedding. 
Figure \ref{method} (a) shows the calculation of distances between cooking-instruction texts used by the proposed method.
Note that both $D_0$ and $D_1$ denote ``Cut a carrot'', whereas $D_2$ denotes ``Boil a potato'' in Figure \ref{method}.
Note also that ``carrot'' is written in Chinese characters in $D_0$ but is written in Hiragana in $D_1$.

In the WMD-based method (see Figure \ref{method} (b)), the sum of the replacement costs for words is defined as the distance between the documents.
The replacement costs for words are given by the cosine distance between the word embeddings.
Our method defines the distance between the cooking-instruction texts as the sum of the replacement costs for the character 3-grams. 
Using our approach, we consider that near-duplicate recipes can be detected robustly, even when paraphrasing, rewriting, typographical errors, and omissions are present in the recipe texts.

\begin{table*}
\centering
\small
\begin{tabular}{|c|c|c|c|c|c|}
\hline
\multicolumn{2}{|c|}{Method} & Near-dup. & Non-dup. A & Non-dup. B & Non-dup. C \\ \hline \hline
\multirow{2}{*}{Character 3-gram} & SGNS & 46 (4.17\%) & 424 (38.41\%) & 331 (29.98\%) & 303 (27.45\%) \\ \cline{2-6}
& fastText & \textbf{47} (4.38\%) & 414 (38.55\%)  & 301 (28.03\%)  & 312 (29.05\%) \\ \hline 
\multirow{2}{*}{Word} & SGNS & \textbf{47} (3.49\%) & 470 (34.89\%) & 382 (28.36\%) & 448 (33.26\%) \\ \cline{2-6} 
& fastText & 46 (4.38\%) & 410 (39.05\%) & 281 (26.76\%) & 313 (29.81\%) \\ \hline \hline
\multicolumn{2}{|c|}{Baseline} & 45 (2.87\%) & 468 (29.89\%) & 477 (30.46\%) & 576 (36.78\%) \\ \hline
\end{tabular}
\caption{Recipe annotation results (``dup.'' denotes ``duplicate'')}
\label{result}
\end{table*}

\begin{comment}
\begin{table*}
\centering
\small
\begin{tabular}{|c|c|c|c|c|c|}
\hline
\multicolumn{2}{|c|}{Method} & Near-dup. & Non-dup. A & Non-dup. B & Non-dup. C \\ \hline \hline
\multirow{2}{*}{SGNS} & character 3-gram & 46 & 424 & 331 & 303 \\ \cline{2-6}
& word & \textbf{47} & 470 & 382 & 448 \\ \hline 
\multirow{2}{*}{fastText} & character 3-gram & \textbf{47} & 414  & 301  & 312 \\ \cline{2-6} 
& word & 46 & 410 & 281 & 313 \\ \hline
\end{tabular}
\caption{Recipe annotation results (``dup.'' denotes ``duplicate'')}
\label{result}
\end{table*}
\end{comment}

\begin{comment}
\begin{table*}
\centering
\small
\begin{tabular}{|c|c|c|c|c|c|}
\hline
\multicolumn{2}{|c|}{Method} & Near-dup. & Non-dup. A & Non-dup. B & Non-dup. C \\ \hline \hline
\multirow{2}{*}{SGNS} & character 3-gram & 46 (4.17\%) & 424 (38.41\%) & 331 (29.98\%) & 303 (27.45\%) \\ \cline{2-6}
& word & \textbf{47} (3.49\%) & 470 (34.89\%) & 382 (28.36\%) & 448 (33.26\%) \\ \hline 
\multirow{2}{*}{fastText} & character 3-gram & \textbf{47} (4.38\%) & 414 (38.55\%)  & 301 (28.03\%)  & 312 (29.05\%) \\ \cline{2-6} 
& word & 46 (4.38\%) & 410 (39.05\%) & 281 (26.76\%) & 313 (29.81\%) \\ \hline
\end{tabular}
\caption{Recipe annotation results (``dup.'' denotes ``duplicate'')}
\label{result}
\end{table*}
\end{comment}

\section{Experiment 1: Investigation of the effectiveness of the proposed method}\label{chap4}

We now describe our experimental evaluation of methods for near-duplicate recipe detection.
Having annotated the candidate near-duplicate recipe pairs, we evaluated methods for near-duplicate recipe detection in terms of the number of detected near-duplicates.

\subsection{Experimental dataset}

We used the Rakuten Recipe dataset, which contained 1,214,650 recipes.
We split the dataset into two datasets, based on the date of publication.
We used 1,210,612 recipes (from June 30, 2010 to October 31, 2016) as the training dataset and used the remaining 4,038 recipes (from November 1, 2016 to November 8, 2016) as the test dataset.

\subsection{Methods}

We utilized the proposed method and a comparison method to extract near-duplicate recipes (see \textsection\ref{4-2-1}).
First, using the training data, we used two embedding-learning algorithms, the Skip-Gram model with Negative Sampling (SGNS) \cite{mikolov:2013} and fastText \cite{bojanowski:2017}, to learn both word embedding and character 3-gram embedding.
We set the hyperparameter values as follows. Both the word and character 3-gram embedding had 100 dimensions, with the window size set to 15. 
%In addition, we set the minimum count of words to 1 for word embedding and to 10 for character 3-gram embedding.

Next, using each recipe in the test dataset as a query, we extracted the top 10 recipes based on the distances between the cooking-instruction texts in the training dataset.
Note that the distances between the calculated cooking-instruction texts differed according to the embedding model used.
We then extracted target recipe pairs for annotation from among the candidate near-duplicate recipe pairs based on ingredients distance (see \textsection\ref{4-2-2}).
%This process identified 1,983 different recipe pairs as target recipe pairs for annotation.

Finally, we annotated these target recipe pairs based on annotation criteria (see \textsection\ref{4-2-3}) and evaluated the proposed method and the comparison method.

\subsubsection{Comparison method}\label{4-2-1}

We adopted a baseline method that extracted candidates of near-duplicate recipe pairs by $tf \cdot idf$ based cosine similarity.
We also extracted target recipe pairs for annotation based on the distances between the cooking-instruction texts, and ingredients distance.

%We adopted the WMD method proposed by Kusner et al. as the comparison method. %(see Figure \ref{method} (b)).
In addition, we also adopted the WMD method as the comparison method. %(see Figure \ref{method} (b)).
By using WMD as the comparison method, we could compare the effectiveness of the proposed method to that of a comparable existing method.

%In addition, we adopted a method that extracts candidate of near-duplicate recipe pairs based on $tf \cdot idf$ vector-based cosine similarity as the baseline method.
%We also extracted target recipe pairs for annotation based on the distances between the cooking-instruction texts and ingredient distance.
%In addition, we adopted a baseline method that extracted candidate of near-duplicate recipe pairs by $tf \cdot idf$ based cosine similarity.
%We also extracted target recipe pairs for annotation based on the distances between the cooking-instruction texts, and ingredient distance.

\subsubsection{Ingredients distance}\label{4-2-2}

Among the ingredients used in recipes, there are ingredients whose description can be paraphrased or that have variations in naming. 
Therefore, we calculated the ingredients distance between the ingredient sets of a candidate original recipe and a candidate near-duplicate recipe by the following process.

First, delete characters in parentheses and symbols from both ingredient lists.
Next, convert ingredient names in both recipes to Katakana and delete ingredients common to both recipes from both ingredient lists.
Then, search for similar words for each ingredient in the candidate near-duplicate recipes based on word embedding. At this stage, if any of the top 3 search results for similar words contains an ingredient of a candidate original recipe, consider it as the same ingredient and delete it from the ingredient lists for both recipes.
Finally, consider the count of ingredients in the ingredient lists of both recipes as the ingredients distance.
Note that, in this study, whenever the ingredients distance was 2 or less, the recipes were considered as annotation targets.

\begin{table*}[htb]
\centering
%\caption{Near-duplicate recipe classification result using machine learning algorithm}
%\label{classfication}
\small
\scalebox{0.8}[0.9]{
%\begin{tabular}{|c|c|c|c|c|c|c|c|c|c|c|c|c|c|c|c|c|}
\begin{tabular}{|c|c|c|c|c|c|c|c|c|c|c|c|c|c|}
\hline
\multicolumn{2}{|c|}{\multirow{2}{*}{Method}} & \multicolumn{3}{c|}{Logistic Regression} & \multicolumn{3}{c|}{SVM (linear)} & \multicolumn{3}{c|}{SVM (RBF)} & \multicolumn{3}{c|}{Random Forest} \\ \cline{3-14}
\multicolumn{2}{|c|}{} & \multicolumn{1}{l|}{F1} & \multicolumn{1}{l|}{Recall} & \multicolumn{1}{l|}{Precision} & \multicolumn{1}{l|}{F1} & \multicolumn{1}{l|}{Recall} & \multicolumn{1}{l|}{Precision} & \multicolumn{1}{l|}{F1} & \multicolumn{1}{l|}{Recall} & \multicolumn{1}{l|}{Precision} & \multicolumn{1}{l|}{F1} & \multicolumn{1}{l|}{Recall} & \multicolumn{1}{l|}{Precision} \\ \hline \hline
\multirow{2}{*}{Character 3-gram} & SGNS & \textbf{0.81} & 0.89 & 0.74 & \textbf{0.81} & 0.97 & 0.70 & \textbf{0.81} & 0.89 & 0.74 & \textbf{0.81} & 0.97 & 0.79 \\ \cline{2-14}
 & fastText & 0.77 & 0.91 & 0.66 & 0.77 & 0.91 & 0.66 & 0.77 & 0.91 & 0.66 & \textbf{0.83} & 0.90 & 0.77 \\ \hline
 \multirow{2}{*}{Word} & SGNS & \textbf{0.81} & 0.68 & 0.91 & \textbf{0.80} & 0.89 & 0.72 & \textbf{0.81} & 1.00 & 0.68 & \textbf{0.85} & 0.93 & 0.79 \\ \cline{2-14}
 & fastText & 0.75 & 0.94 & 0.63 & 0.75 & 0.88 & 0.65 & 0.76 & 0.97 & 0.63 & \textbf{0.89} & 0.97 & 0.83 \\ \hline
\end{tabular}
}
\caption{Near-duplicate recipe classification results using machine learning algorithms}
\label{classfication}
\end{table*}

\subsubsection{Annotation criteria}\label{4-2-3}

We annotated the recipe pairs using four labels: near-duplicate, non-duplicate A, non-duplicate B, and non-duplicate C.
The main annotation criteria are as follows.

\textbullet Near-duplicate: ingredients are exactly the same and cooking-instruction texts are also the same except for additional expressions. %(e.g., at the end of sentences).

\textbullet Non-duplicate A: the same dish, with some common ingredients and common cooking-instruction texts.

\textbullet Non-duplicate B: different main ingredients, but with common cooking-instruction texts (except for the handling of the different ingredients).

\textbullet Non-duplicate C: different cooking methods, different cooking-instruction texts, and different ingredients.

The first author and another participant in the experiment both annotated the same 300 recipe pairs using these annotation criteria. We then applied Cohen's $\kappa$ coefficient \cite{cohen:1960} to assess the degree of agreement between the two sets of annotations. The resulting $\kappa$ value of 0.903 indicated almost perfect agreement \cite{landis:1977}.
This was regarded as confirming the reliability of the annotation criteria and the first author completed the annotation of the remaining recipe pairs. 

\subsection{Experimental results}

%Table \ref{result} shows the results of the annotation, indicating that the non-duplicate A class contained the highest proportion of annotated recipe pairs.
%This result implies that there are many low-quality recipes that copy a cooking-instruction text from other recipes and merely replace the ingredient list. 
Table \ref{result} shows the results of the annotation.
The percentage of recipe pairs that were labeled ``near-duplicate'' were 3.49\% or more when using the proposed method (and the comparison method), whereas it was 2.87\% when using the baseline method.
%The percentage of recipe pairs that were labeled ``near-duplicate'' was 2.87\% when using the baseline method, whereas they were 3.49\% or more when using the proposed method and the comparison method.
%In contrast, the percentage of recipe pairs that were labeled ``near-duplicate'' was 3.49\% or more when using the proposed method and the comparison method.
%Besides, the percentage of recipe pairs that were labeled ``non-duplicate C'' was especially many when using the baseline method.
This result indicates that the proposed method using the character 3-gram embedding and the comparison method using the word embedding are superior to the baseline method.

Based on the number of detected near-duplicate recipes, there was no quantitative difference between the proposed method and the comparison method; however, as a result of qualitative analysis, the following differences were found between detected near-duplicate recipes by the proposed method and those detected by the comparison method.

\noindent
\textbf{Strengths.} The near-duplicate recipe pairs detected only by the proposed method had the following features. One recipe contained typographical errors and one recipe replaced materials by abbreviations (e.g., ``Cucumber and tomato ...'' by ``A. ...''). Moreover, the word-based distances are large for long words because of morphological analysis problems, even if the words are semantically similar. 

\noindent
\textbf{Weaknesses.} The near-duplicate recipe pairs detected only by the comparison method had the following features. The word order was changed in one recipe, phrases were rewritten in another recipe, and expressions were paraphrased to give similar expressions in sentence units.

\section{Experiment 2: Investigation of the effectiveness of embedding-learning algorithms}

We now describe our experiment for evaluating the calculation accuracy of the distance between cooking-instruction texts using machine learning algorithms.

\subsection{Methods}

We used the near-duplicate recipe pairs as positive examples and the remaining recipe pairs as negative examples. 
We classified the near-duplicate recipe candidates based on the distance between cooking-instruction texts and ingredients distance by following four machine learning algorithms: Logistic Regression, Support Vector Machine (Linear kernel), Support Vector Machine (Radial Basis Function (RBF) kernel \cite{buhmann:2003}), and Random Forest.
The experimental data included approximately 50 recipe pairs as positive examples (near-duplicates) and approximately 1,000 recipe pairs as negative examples (the remaining pairs).
It is well known that unbalanced data can affect the results of binary classification \cite{yen:2006}.
To avoid such potential problems, we applied undersampling to extract features from the negative examples at random to match the number of positive examples. 
We utilized leave-one-out cross-validation, grid-search, and tuned hyperparameters in the evaluation.

\subsection{Experimental results}

Table \ref{classfication} shows the results of the classification.
Note that F1 scores of at least 0.7 were obtained for all machine learning algorithms.
Moreover, in most cases, Recall was greater than Precision. 
These results indicate that the number of non-near-duplicate recipes that were classified wrongly is larger than the number of near-duplicate recipes that were classified wrongly, which is a good result because it is the presence of near-duplicate recipes that represents the greater inconvenience for the recipe websites. 

Analyzing the classification results, we see that most of the non-near-duplicate recipes that were classified wrongly had been labeled as ``non-duplicate A''.
As described in the annotation criteria, recipe pairs labeled non-duplicate A have the following features: the same dish, with some common ingredients, and with cooking-instruction texts in common.
This criterion differs only slightly from that for near-duplicates and is therefore the most likely reason for the classification errors.
However, near-duplicate recipes that were classified wrongly had a greater distance between their cooking-instruction texts and/or a greater number of ingredient similarities than other near-duplicate recipes. This would explain the classification errors.

\section{Conclusion}

We have proposed a method that extends WMD to character 3-gram embedding, enabling the Character 3-gram Mover's Distance to be utilized.
Our experiments demonstrated that the proposed method can detect near-duplicate recipes containing typographical errors and omissions that cannot be detected by a comparison method.
However, among the near-duplicate recipes detected by the comparison method were some that were not detected by the proposed method.

In future research, we plan to improve the algorithm for detecting near-duplicate recipes.
In particular, we are considering combining WMD with the Character 3-gram Mover's Distance and the use of a language-representation model (e.g., ELMo \cite{peters:2018} or BERT \cite{devlin:2019}).

%\bibliography{emnlp-ijcnlp-2019}
\bibliography{main}
\bibliographystyle{acl_natbib_nourl}

\end{document}